\documentclass{elsart}

\usepackage{natbib}

 \usepackage{epsfig}

\begin{document}

\begin{frontmatter}



\title{Relationship between pulse width and energy in GRB 060124:
 from X-ray to gamma-ray bands}


\author[a,b]{Fu-Wen Zhang},
\ead{fwzhang@ynao.ac.cn}
\author[c,d]{Yi-Ping Qin}
\ead{ypqin@ynao.ac.cn}
\address[a]{National Astronomical Observatories/Yunnan
Observatory, Chinese Academy of Sciences, P.O. Box 110, Kunming,
Yunnan 650011, China}
\address[b]{The Graduate School of the Chinese Academy of
Sciences, P.O. Box 3908, Beijing 100039, China}
\address[c]{Center for Astrophysics, Guangzhou University,
Guangzhou 510006, China}
\address[d]{Physics Department, Guangxi University, Nanning,
 Guangxi 530004, China}

\begin{abstract}
GRB 060124 is the first event that both prompt and afterglow
emission were observed simultaneously by the three \emph{Swift}
instruments. Its main peak also triggered Konus-Wind and HETE-II.
Therefore, investigation on both the temporal and spectral
properties of the prompt emission can be extended to X-ray bands. We
perform a detailed analysis on the two well identified pulses of
this burst, and find that the pulses are narrower at higher
energies, and both X-rays and gamma-rays follow the same $w - E$
relation for an individual pulse. However, there is no a universal
power-law index of the $w - E$ relation among pulses. We find also
that the rise-to-decay ratio $r/d$ seems not to evolve with $E$ and
the $r/d$ values are well consistent with that observed in typical
GRBs. The broadband spectral energy distribution also suggest that
the X-rays are consistent with the spectral behavior of the
gamma-rays. These results indicates that the X-ray emission tracks
the gamma-ray emission and the emissions in the two energy bands are
likely to be originated from the same physical mechanism.
\end{abstract}


\end{frontmatter}

% main text
\section{Introduction}
The \emph{Swift} satellite (Gehrels et al., 2004) was successfully
launched on 20th November 2004. It is a multi-wavelength
observatory, covering the gamma-ray, X-ray and UV/optical bands.
Thanks to its rapid repointing capability, the mission has
revolutionized the Gamma-Ray Burst (GRB) observations in many
aspects (for recent reviews, see M\'{e}sz\'{a}ros, 2006; Fox and
M\'{e}sz\'{a}ros, 2006; Zhang, 2007). The prompt slewing capability
of the X-Ray Telescope (XRT, Burrows et al., 2005) and UV-Optical
Telescope (UVOT, Roming et al., 2005) allows the satellite to
swiftly catch very early X-ray and UV/optical signals following the
GRB prompt emission detected by the Burst Alert Telescope (BAT,
Barthelmy et al., 2005).

In the pre-\emph{Swift} era, the temporal and spectral behaviors of
GRB prompt emission have been studied extensively. It is found that
a pulse of the prompt gamma-rays at lower energy bands tend to be
wider, which is roughly depicted as $w\propto E^{-0.4}$ ($w - E$
relation; Fishman et al., 1992; Link et al., 1993; Fenimore et al.,
1995; Norris et al., 1996, 2005; Piro et al., 1998; Costa, 1999;
Nemiroff, 2000; Feroci et al., 2001; Crew et al., 2003; Peng et al.,
2006; Zhang et al., 2007b). The photons at lower energy bands also
lag behind that of the photons at higher energy bands (the so-called
spectral lag; Cheng et al., 1995; Norris et al., 1996, 2005; Norris,
Marani and Bonnell, 2000; Wu and Fenimore, 2000; Chen et al., 2005;
Yi et al., 2006; Zhang et al., 2006b, 2006c; Peng et al., 2007).

Although many attempts have been made to explain the $w - E$
relation and the spectral lag behavior(e.g. Fenimore et al. 1995;
Cohen et al., 1997; Chiang, 1998; Dermer, 1998; Kazanas et al. 1998;
Piran, 1999; Wang et al., 2000; Nemiroff, 2000; Qin et al., 2004,
2005; Shen et al., 2005; Lu et al., 2006; Zhang et al., 2007b; Dado,
Dar and De R\'{u}jula, 2007), its nature remains a matter of active
debate in the community. On the other hand, it is unclear whether
this correlation can be extended to X-ray bands. The broadband
observations showed that the X-ray emission of some GRBs have
unusual properties. in't zand et al. (1999) found that the prompt
X-ray emission of GRB 980519, measured by BeppoSAX, undergoes a
strong soft-to-hard-to-soft evolution. An exceptionally intense
gamma-ray burst, GRB 030329, was detected and localized by the
instruments on board the High Energy Transient Explorer satellite
(HETE). It's lightcurve has a distinct, bright, soft X-ray
component(Vanderspek et al. 2004). A thermal emission component is
identified from the XRT data of a nearby XRF 060218 (Campana et al.
2006), but its non-thermal X-rays are from the same emission
component as the gamma-rays (Liang et al., 2006). Vetere et al.
(2006) analyzed the X-ray temporal and spectral characteristics of
ten GRBs detected by the WFCs on board BeppoSAX and argued that
there exist two components (slow and fast) in the X-ray emission.
These facts suggest that the physics of these prompt X-rays are also
very uncertain.

With benefit from the Swift satellite, the prompt X-rays of some
bursts were observed, e.g. GRB 050117 (Hill et al., 2006),
GRB050713A (Morris et al., 2007), GRB050820A (Cenko et al., 2006),
GRB 060124 (Romano et al., 2006), GRB 060218 (Campana et al., 2006;
Liang et al., 2006); GRB 061121 (Page et al., 2007) and GRB 070129(
Godet et al., 2007; Krimm et al., 2007). These panchromatic
observations unveiled the unprecedented spectral and temporal
information of GRB prompt emission. This also makes it possible to
measure their temporal structures and to examine whether the
well-known $w - E$ relation can be extended to X-ray bands for these
bursts. Except for GRB 060124 and 060218, these bursts have very
complicated temporal structures that consist of a series of
overlapping pulses. Liang et al. (2006) analyzed the non-thermal
emission of GRB 060218 from the gamma-ray to X-ray bands and
obtained that $w\propto E^{-0.31\pm0.03}$, which roughly satisfies
the $w- E$ relation and the relation between spectral lag and
luminosity derived from typical GRBs (Fenimore et al. 1995; Norris
et al., 1996, 2000, 2005), although it has the longest pulse
duration and spectral lag observed to date among the observed GRBs.
They suggested that the prompt X-rays and gamma-rays are from the
same component. In this paper we present a detailed analysis on the
prompt emission of GRB 060124 to examine whether the $w-E$ relation
can be extended to X-rays. In Section 2, we present the data
reduction. Results are given in Section 3. Our conclusions are
presented in Section 4.

\section{Data Reduction}
GRB 060124 was detected by \emph{Swift}-BAT at 15:54:52 UT on 24
January 2006 (trigger 178750, Holland et al., 2006; Fenimore er al.,
2006), located at right ascension $05^{h}08^{m}10^{s}$ and
declination $+69^{0}42^{'}33^{''}$, with an uncertainty of $3^{'}$.
The burst also triggered Konus-Wind (Golenetskii et al., 2006)
559.351 s and the FREGATE instrument aboard HETE-II (HETE trigger
4012, Lamb et al., 2006) 557.7 s after the BAT trigger. The prompt
emission was observed simultaneously by XRT ($T_{0}+104$ s, where
$T_{0}$ denotes the BAT trigger time) and UVOT at $V=16.96\pm0.08$
($T_{0}+183$ s) and $V=16.79\pm0.04$ ($T_{0}+633$ s).
\emph{Swift}-BAT first triggered on a precursor, then after about
500 s three major peaks following the precursor were observed. The
burst is one of the longest GRBs (even excluding the precursor)
recorded by either BATSE or \emph{Swift}, and it is the first event
that three \emph{Swift} instruments have a clear detection of both
the prompt and the afterglow emission (Romano et al., 2006).

The main peak data of GRB 060124 are not included in the event data
since BAT was triggered by the precursor of the burst and the event
data lasted only $t\sim 300$ s after the trigger. The main peaks of
the burst are visible in the mask-tagged 4 channel (15-25, 25-50,
50-100 and 100-350 keV bands) light curves, where %the integration
the time bin is 1.6 s, generated by the flight software (see Fig.
1). The XRT observations in sequence 000 included a %total
net exposure time of $t\sim 867$ s in Windowed Timing (WT) mode
and 15 s in Photon Counting (PC) mode. The WT mode data recorded
the main burst and were adopted in this work. The data were
reduced with the standard XRTDAS tools, using the latest
calibration files available in CALDB and the standard data
screening. For our analysis, events in the 0.2-10 keV band with
grades 0-2 were used (see Burrows et al. 2005). The data of the WT
mode during sequence 000 were affected by pile-up. To account for
this effect, we extract a rectangular $40\times20$-pixel region
with a $4\times20$-pixel region excluded from its center as the
source region. The background region is a box ($40\times20$-pixel)
far away from the source. The XRT-WT events were extracted in
three energy bands, 0.2-1, 1-4 and 4-10 keV (the corresponding
backgrounds were subtracted). These are also shown in Fig. 1. The
count rates of the burst were also recorded by Konus in three
energy ranges: G1 (18-70 keV), G2 (70-300 keV), and G3 (300-1160
keV). The Konus trigger data are recorded from $T_{KW}-$0.512 s to
$T_{KW}$+229.632 s with time resolution ranging from 2 to 256 ms.
The data before $T_{KW}-$0.512 s are collected in the waiting mode
with 2.944 s time resolution. The Konus time history of this burst
in the G3 band (2.944 s time resolution) is reported in Fig. 1 as
well.

\section{Results}

As shown in Fig. 1, the largest peak ($\sim$560-580 s) exists
significantly in all eight energy bands. The temporal properties of
this peak in both the X-ray and gamma-ray bands are worth
discussing. This peak consists of three main pulses. We focus on two
apparent pulses among them.

Kocevski et al. (2003) developed an empirical expression, which
can be used to fit the pulses of GRBs. This function is written as
\begin{equation}
F(t)=F_{m}(\frac{t+t_{0}}{t_{m}+t_{0}})^{r}[\frac{d}{d+r}+\frac{r}{d+r}
(\frac{t+t_{0}}{t_{m}+t_{0}})^{(r+1)}]^{-\frac{r+d}{r+1}},
\end{equation}
where $t_{m}$ is the time of the maximum flux ($F_{m}$) of the
pulse, $t_{0}$ is the offset time, $r$ and $d$ are the rising and
decaying power-law indices, respectively.

From the BAT light curves we find that the peak considered here is
apparently separable at about $T_{0}+540$ s, but it is not
separable in the three XRT energy ranges due to overlapping.
Norris et al. (1996) developed a method to deconvolve an
overlapped GRB temporal profile into pulses. We use this method
deconvolve the temporal profiles of the burst from 450 s to 540 s
into seven pulses in the three XRT energy bands, where Eq. 1 is
adopted as the pulse model. An interactive graphical IDL routine
and a least-squares algorithm are applied to the pulse fit. The
front pulses obtained from the fit are subtracted from the XRT
light curves, which yields new data of the light curves. In this
way, the peaks after 540 s are no more affected by the front
overlapping pulses. Presented in Fig. 2 are these new XRT light
curves and the BAT light curves.

It is known that the burst temporal profiles are self-similar
across energy bands (e.g. Norris et al. 1996). We deconvolve the
largest peaks from 540 s to 650 s into three pulses for all XRT
and BAT energy bands with the same methods described above, where
the new XRT data are adopted. The fitting results are shown in
Fig. 2 as well\footnote{Due to the heavily overlapping in 0.2-1
keV and 1-4 keV bands, we cannot obtain very robust data within
these bands from this analysis. Thus, parameters derived from
these two bands are regarded as merely qualitative results.}. We
label the two apparent pulses in the concerned energy bands as
Pulse 1 (that with the smaller magnitude) and Pulse 2 (that with
the larger magnitude), respectively (see Fig. 2).

The data recorded by Konus before $T_{KW}-$0.512 s are collected
in the waiting mode with 2.944 s time resolution. Since this time
resolution is low, the corresponding data cannot be used in our
analysis. We adopt only the 256 ms trigger data after
$T_{KW}-$0.512 s (558.839 s after the BAT trigger) to preform the
temporal analysis in the 300-1160 keV band. The data are also
displayed in Fig. 2. We find from the figure that, in the 300-1160
keV band, only one pulse (Pulse 2) can be identified and can be
fitted with Eq. 1.

We measure the pulse width $w$, and the ratio of the rising time
to the decaying time $r/d$ at the full-width half-maximum (FWHM)
of the fitting curves. The uncertainties of these quantities are
derived from the errors of the fitting parameters according to the
error transform function. The results are listed in Table 1, where
the errors are reported in the $1\sigma$ confidence level.

There exists a significant trend that pulses are narrower at higher
energies. The $w- E$ relations for two pulses are presented in Fig.
3,  where $E$ is the geometric mean of the lower and upper
boundaries of the energy band. It shows that the two quantities are
correlated. Following Fenimore et al. (1995), we parameterize the
dependence of the pulse width on energy by a power law. The best fit
yields $w\propto E^{-0.47\pm0.05}$ (N=5) for Pulse 1 and $w\propto
E^{-0.23\pm0.03}$ (N=6) for Pulse 2. Please note that the data in
0.2-1 and 1-4 keV bands are not included in the fits. It is found
that the power-law index for Pulse 1 is roughly consistent with that
previously observed in typical GRBs, but it is much shallower for
Pulse 2, similar to that observed in GRB 060218 (Liang et al.,
2006). Note that the distribution of the index for a typical GRB
sample has a large dispersion, with a median of $\sim$ -0.4 (see,
Jia and Qin, 2005; Peng et al., 2006; Zhang et al., 2007b). Thus, it
is possible that there is no a universal power-law index of the $w -
E$ relation. For these two pulses, we find that their $r/d$ values
are well consistent with that observed in typical GRBs (e.g., Norris
et al., 1996, 2005; Liang et al., 2002; Peng et al., 2006), but no
apparent relationship between $r/d$ and $E$ is observed.

Broadband spectral energy distribution (SED) is also helpful to
discriminate different emission components. We derive the SED from
the peak fluxes of an individual pulses at different energy bands by
using the spectral data\footnote{Due to the spectral data observed
by Konus-Wind are not provided, the peak flux of Pulse 2 in 300-1160
keV is not measured.}. The results are listed in Table 2 (also see
Fig. 4). From Table 2 and Fig. 4, one can find that the peak fluxes
for the two pulses become larger at higher energies, suggesting that
the X-rays are the same emission component of the gamma-rays.

\section{Conclusions and Discussion}

We have analyzed the temporal properties of the two well-identified
pulses of GRB 060124 from X-ray (0.2-10 keV) to gamma-ray (15-1160
keV) energy bands. We find that the pulse width $w$ is
energy-dependent for the two pulses in eight energy bands (0.2-1,
1-4, 4-10, 15-25, 25-50, 50-100, 100-350 and 300-1160 keV). The
pulses are found narrower at higher energies, and both X-rays and
gamma-rays follow the same $w - E$ relation for an individual pulse.
However, we find that there is no a universal power-law index of the
$w - E$ relation among pulses. We also find that the rise-to-decay
ratio $r/d$ seems not to evolve with $E$ and the $r/d$ values are
well consistent with that observed in typical GRBs. The peak fluxes
for the two pulses become larger at higher energies. These results
indicates that the X-ray emission tracks the gamma-ray emission and
the emissions in the two energy bands are likely to be originated
from the same physical mechanism.

One remarkable advance from {\em Swift} is that the on-board XRT has
established a large sample of X-ray lightcurves from tens of seconds
to days (Zhang et al., 2006a; Nousek et al., 2006; O'Brien et al.,
2006). The physical mechanisms of these X-rays are of great
uncertain and is on debate in the GRB community (see review by
Zhang, 2007). It is possible that the mechanisms are diverse (e.g.,
Zhang et al., 2006a; Zhang et al., 2007a, Liang et al., 2007, 2008).
As we show here that some X-rays are possibly from the same emission
component (see also Liang et al., 2006 for GRB 060218). However,
Vetere et al. (2006) analyzed the temporal and spectral
characteristics of X-rays for ten GRBs detected by the WFCs on board
BeppoSAX and argued that there exist two components (slow and fast)
in the X-ray emission. This feature actually is also seen in GRB
030329 (Vanderspek et al., 2004). Both temporal and spectral
properties are critical to discriminate these components.

\textbf{Acknowledgments}

We are grateful to Dr. Valentin Pal'shin for providing the
Konus-Wind data. We should also like to thank the anonymous
referee for helpful suggestions and En-Wei Liang, Jin-Ming Bai and
Bin-Bin Zhang for valuable discussion. This work is supported by
National Natural Science Foundation of China (No. 10573030, No.
10533050 and No. 10463001).

% \appendix

% \section{}
% \label{}

% Bibliographic references with the natbib package:
% Parenthetical: \citep{Bai92} produces (Bailyn 1992).
% Textual: \citet{Bai95} produces Bailyn et al. (1995).
% An affix and part of a reference:
%   \citep[e.g.][Ch. 2]{Bar76}
%   produces (e.g. Barnes et al. 1976, Ch. 2).

\clearpage

\begin{figure}
\begin{center}
\includegraphics*[width=15cm]{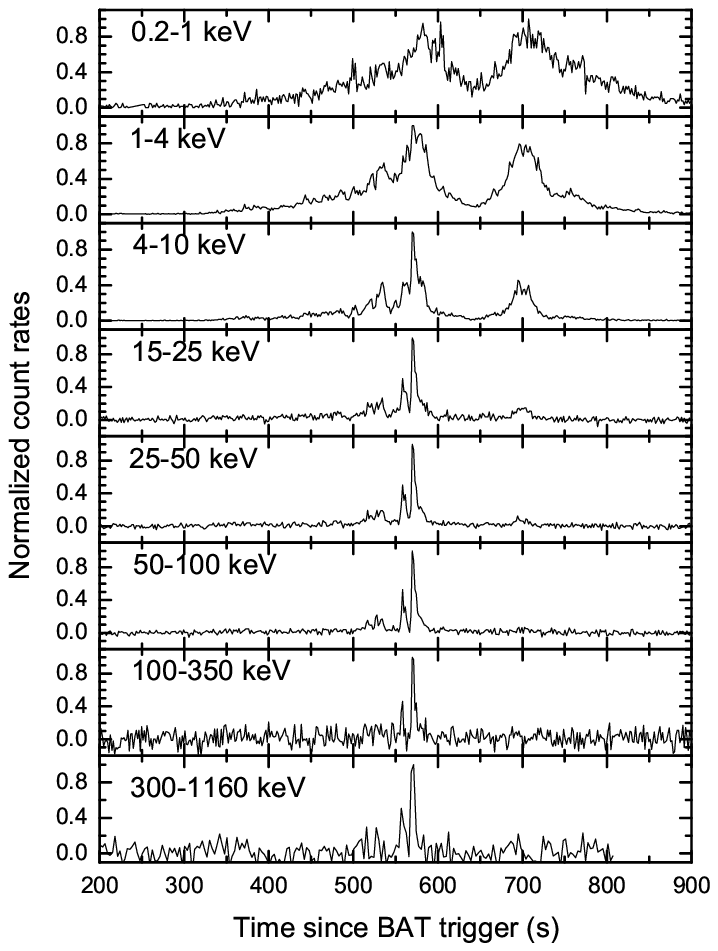}
\end{center}
\caption{XRT, BAT and Konus light curves of the prompt emission of
GRB 060124. The count rates have been normalized to the peak of
each light curve.\label{fig1}}
\end{figure}

\begin{figure}
\begin{center}
\includegraphics*[width=15cm]{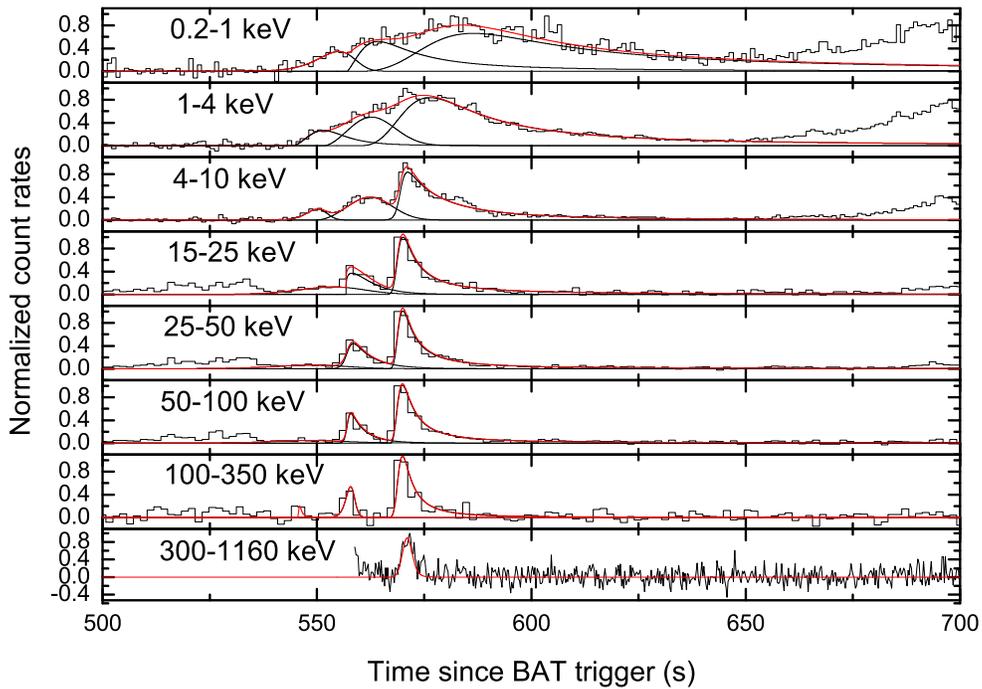}
\end{center}
\caption{XRT, BAT and Konus light curves around the largest flux
of the burst, where the effects on the XRT light curves after 540
s by the front overlapping pulses have been checked. The three
fitting pulses accounting for the largest peak (540-650 s) are
also plotted in this figure. The red lines represent the
superpositions of the three fitting pulses (in the 300-1160 keV
band, only one fitting pulse is adopted). \label{fig2}}
\end{figure}

\begin{figure}
\begin{center}
\includegraphics*[width=15cm]{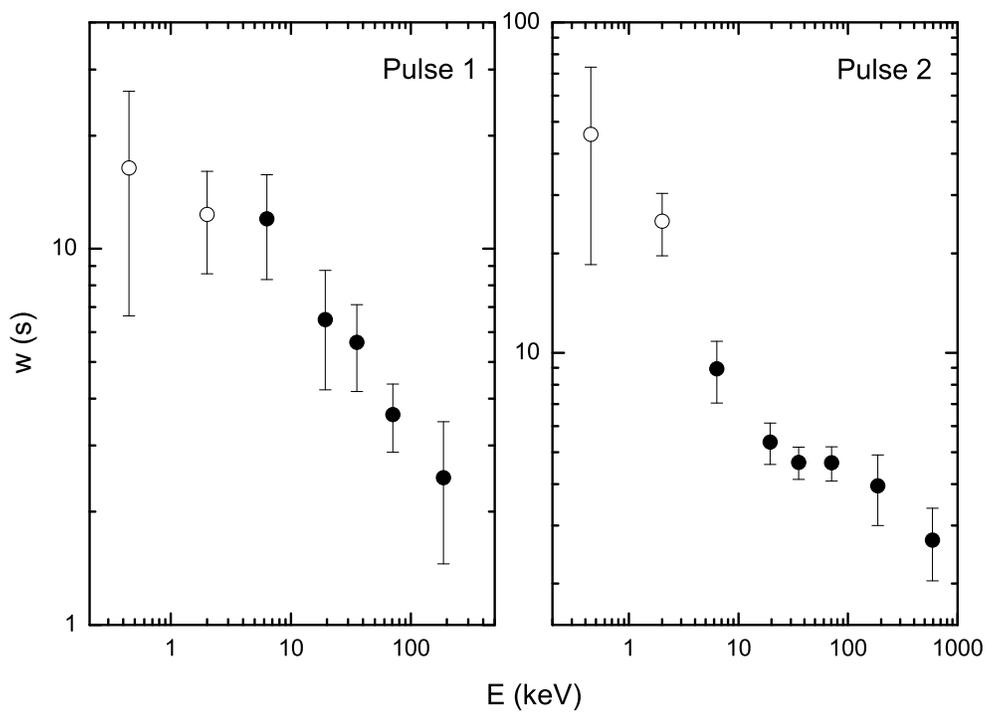}
\end{center}
\caption{Relationship between $w$ and $E$, where open circles
denote $w$ estimated in 0.2-1 and 1-4 keV bands, and filled
circles denote $w$ in other energy bands. \label{fig2}}
\end{figure}

\begin{figure}
\begin{center}
\includegraphics*[width=15cm]{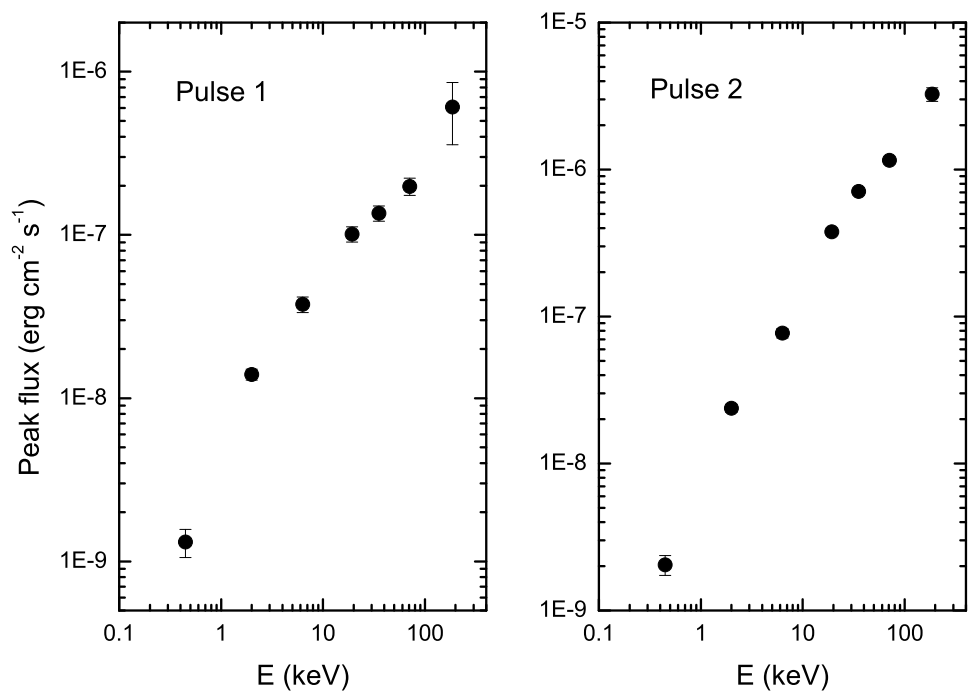}
\end{center}
\caption{Relationship between the pulse peak flux and energy.
\label{fig2}}
\end{figure}

\clearpage

\begin{table}[htb]
\center \caption{Multi-wavelength temporal characteristics of the
two pulses in GRB 060124.} \label{table:1}
\newcommand{\m}{\hphantom{$-$}}
\newcommand{\cc}[1]{\multicolumn{1}{c}{#1}}
\renewcommand{\tabcolsep}{2pc} % enlarge column spacing
\renewcommand{\arraystretch}{1.2} % enlarge line spacing
\begin{tabular}{@{}lcccc}
\hline\hline
Band (keV)  & Pulse 1 & &Pulse 2 & \\
  & w (s) & r/d  &  w (s) & r/d  \\
\hline
        0.2-1   &16.4$\pm$9.8 &0.41$\pm$0.17 &45.8$\pm$27.4&0.40$\pm$0.10\\
        1-4     &12.3$\pm$3.7&0.50$\pm$0.31 &25.0$\pm$5.3 &0.45$\pm$0.09 \\
        4-10    &12.0$\pm$3.7&0.48$\pm$0.34 &8.9$\pm$1.9  &0.27$\pm$0.09 \\
        15-25   &6.5$\pm$2.3 &0.26$\pm$0.15 &5.4$\pm$0.8  &0.38$\pm$0.14 \\
        25-50   &5.6$\pm$1.5 &0.44$\pm$0.20 &4.7$\pm$0.5  &0.40$\pm$0.14 \\
        50-100  &3.6$\pm$0.7 &0.36$\pm$0.20 &4.6$\pm$0.5  &0.42$\pm$0.15 \\
        100-350 &2.5$\pm$1.0 &0.49$\pm$0.32 &4.0$\pm$1.0  &0.47$\pm$0.26 \\
        300-1160&...         &...           &2.7$\pm$0.7  &0.48$\pm$0.24 \\
\hline
\end{tabular}\\[2pt]
\end{table}

\begin{table}[htb]
\center \caption{Peak fluxes of the individual pulses at different
energy bands in GRB 060124.}

\begin{tabular}{ccc}
\hline\hline
  & Pulse 1 &Pulse 2 \\
Energy band   & Peak energy flux  & Peak energy flux \\
(keV) &(erg cm$^{-2}$ s$^{-1}$) &(erg cm$^{-2}$ s$^{-1}$) \\

 \hline
        0.2-1   &(1.3$\pm0.3)\times10^{-9}$  &(2.1$\pm0.3)\times10^{-9}$ \\
        1-4     &(1.4$\pm0.1)\times10^{-8}$  &(2.4$\pm0.1)\times10^{-8}$ \\
        4-10    &(4.8$\pm0.4)\times10^{-8}$  &(7.7$\pm0.6)\times10^{-8}$ \\
        15-25   &(1.0$\pm0.1)\times10^{-7}$  &(3.8$\pm0.2)\times10^{-7}$ \\
        25-50   &(1.4$\pm0.1)\times10^{-7}$  &(7.1$\pm0.2)\times10^{-7}$ \\
        50-100  &(2.0$\pm0.2)\times10^{-7}$  &(1.2$\pm0.1)\times10^{-6}$ \\
        100-350 &(6.1$\pm0.3)\times10^{-7}$  &(3.3$\pm0.3)\times10^{-6}$ \\
\hline
\end{tabular}\\[2pt]
\end{table}

\end{document}